\documentclass[aps,prd,superscriptaddress,showpacs,floatfix,twocolumn]{revtex4-1}
\usepackage{epsfig}
\usepackage[utf8]{inputenc}
\usepackage{float}
\usepackage{graphicx}
\usepackage{amsmath}
\usepackage{verbatim}
\usepackage{xcolor}
\usepackage{caption}
\usepackage{color}
\usepackage{soul}

\begin{document}

\title{Pion partonic distributions in a statistical model from
  pion-induced \\ Drell-Yan and $J/\Psi$ production data}

\author{Claude Bourrely}
\affiliation{Aix Marseille Univ, Universit\'e de Toulon, CNRS, CPT, Marseille,
France}
\author{Wen-Chen Chang}
\affiliation{Institute of Physics, Academia Sinica, Taipei 11529, Taiwan} 
\author{Jen-Chieh Peng}
\affiliation{Department of Physics, University of Illinois at 
Urbana-Champaign, Urbana, Illinois 61801, USA}

\date{\today}

\begin{abstract}
We present a new analysis to extract pion parton distribution 
functions (PDFs) within the framework of the statistical model. Starting from
the statistical model first developed for the spin-1/2 nucleon, we
extend this model to describe the spin-0 pion. Based on a combined fit 
to both the pion-induced Drell-Yan data and the pion-induced
$J/\Psi$ production data, a new
set of pion PDFs has been obtained. The inclusion
of the $J/\Psi$ production data in the combined fit has provided additional
constraints for better determining the 
gluon distribution in the pion. We also compare the pion PDFs obtained in 
the statistical model with other existing pion PDFs.
\end{abstract}
\maketitle
\clearpage

\section{Introduction}

The description of proton parton distribution functions (PDFs) in
terms of a statistical model approach, initiated 20 years
ago~\cite{BBS2002}, has provided pertinent insights on the flavor,
spin, and momentum dependencies of the partonic constituents of the
proton.  An important feature of the statistical model is the natural
connection between the valence and the sea quark distributions through
the correlations in their Fermi-Dirac momentum distributions. The
proton PDFs obtained in the statistical model can describe existing
deep inelastic scattering (DIS) and Drell-Yan (DY) data very well. The
statistical model approach also led to many successful
predictions~\cite{BBS2005}, including the flavor asymmetries of $\bar
d(x) > \bar u(x)$ for the unpolarized
sea~\cite{NA51,Hawker,Peng1998,Towell2002,Dove2021} and $\Delta \bar
u(x) > 0 > \Delta \bar d(x)$ for the polarized
sea~\cite{STAR2014,STAR2020}. An updated overview of the major results
obtained in the statistical model for describing the nucleon parton
distributions can be found in~\cite{BS2015}.

The success of the statistical model for understanding many
characteristics of the proton PDFs naturally suggests the feasibility
of extending this approach to describe the partonic structures of
other hadrons. Of particular importance are the partonic structures of
pion and kaon, which have attracted much attention recently.  As the
lightest quark-antiquark bound state as well as a Goldstone boson due
to the spontaneous breaking of the chiral symmetry, the pion is a
unique hadronic system for exploring nonperturbative aspects of QCD.

The mounting interest in the partonic structure of pion is reflected 
in many recent studies under various theoretical frameworks including 
the chiral-quark 
model~\cite{Nam:2012vm,Watanabe:2016lto, Watanabe:2017pvl}, 
Nambu-Jona-Lasinio model~\cite{Hutauruk:2016sug},
light-front Hamiltonian~\cite{Lan:2019vui, Lan:2019rba}, holographic
QCD~\cite{deTeramond:2018ecg, Watanabe:2019zny}, maximum entropy
method~\cite{Han:2018wsw}, and the continuum functional approach using
Dyson-Schwinger equations (DSE)~\cite{Chang:2014lva, Chang:2014gga,
  Chen:2016sno, Shi:2018mcb, Bednar:2018mtf, Ding:2019lwe}.
Moreover, a major advance in lattice QCD~\cite{Ji:2013dva} has led
several groups to perform calculations of the momentum ($x$) dependence
of pion partonic distributions~\cite{Chen:2018fwa, Sufian:2019bol, 
Izubuchi:2019lyk, Joo:2019bzr, Sufian:2020vzb, Gao:2020ito, Alexandrou:2021mmi,
Fan:2021bcr}. 
While the majority of these theoretical studies is focused on the pion
valence quark distributions, recent DSE approach has also calculated the
sea-quark and gluon distributions of pion~\cite{Ding:2019lwe}.

On the experimental side, new information
on the pion PDFs has been obtained in the COMPASS experiment with
pion-induced dimuon production~\cite{Compass2017,CompassNew}. 
Additional measurements
aiming at improved accuracy are planned for the future 
AMBER experiment at CERN~\cite{AMBER}. 
Another experimental approach for probing pion
PDFs is the tagged deep-inelastic scattering (TDIS) involving
the DIS off the pion cloud via the Sullivan process~\cite{Sullivan}.
The TDIS approach is being pursued at the Jefferson
Laboratory~\cite{Thia} and planned for the future Electron-Ion
Collider (EIC)~\cite{EIC}.
The interest in pion partonic structure has also led to several recent
extractions of pion PDFs via global fits to various existing 
data~\cite{JAM,BS2019,Xfitter,BBP2021,JAM2021}. Until recently,
knowledge of the pion PDFs was limited to
global analyses performed more than two decades ago:
OW~\cite{Owens:1984zj}, ABFKW~\cite{Aurenche:1989sx}, 
SMRS~\cite{Sutton:1991ay}, GRV~\cite{Gluck:1991ey}, and 
GRS~\cite{Gluck:1999xe}. These analyses were based mostly on
pion-induced Drell-Yan and prompt-photon 
production data. New global analyses were performed recently,
using Drell-Yan data in 
BS~\cite{BS2019} as well as both the Drell-Yan and
direct-photon data in xFitter~\cite{Novikov:2020snp}. The analysis of
JAM~\cite{JAM} included both the Drell-Yan data and 
the leading-neutron tagged electroproduction data.

The first extraction of the pion's PDFs based on the statistical model
approach was reported in~\cite{BS2019}. This work was followed by another 
recent analysis performed by BBP~\cite{BBP2021} which significantly 
reduced the number of parameters
by imposing some constraints based on symmetry principles.
This new analysis shows that a good description of the
existing $\pi^-$-induced Drell-Yan data can be obtained in the
statistical model approach with a
reduced number of parameters.

As the pion-induced Drell-Yan cross sections are dominated by
$\bar{q}$-$q$ annihilation, they essentially probe the valence quark
distribution in the pion, but leave the sea and the gluon
distributions largely unconstrained. Two recent
studies~\cite{Chang:2020rdy,Chang2021} showed that the existing
pion-induced $J/\Psi$ production data can impose useful additional
constraints on the pion PDFs. Utilizing the theoretical frameworks of
the color evaporation model (CEM) as well as the non-relativistic QCD
(NRQCD), it was found that existing pion-induced $J/\Psi$ production
data are sensitive to pion gluon distribution at relatively large $x$
region. This result suggests the importance of including existing
pion-induced $J/\Psi$ production data in a global fit to extract the
pion PDFs. In this paper, we present a new extraction of pion PDFs in
the framework of statistical model via a global fit to both the
Drell-Yan and the $J/\Psi$ production data with pion beam.
 
\section{Parametrizations of meson PDFs in the statistical model}

We begin by defining the notations of the various parton distribution functions
for pions. After imposing the particle-antiparticle charge-conjugation (C)
symmetry for the parton distributions in charged pions, we can define the
PDFs of $\pi^+$ and $\pi^-$ as follows:
\begin{equation}
U(x) \equiv  u_{\pi^+}(x) = \bar u_{\pi^-}(x);~~~D(x) \equiv
\bar d_{\pi^+}(x) = d_{\pi^-}(x)~.
\label{eq1}
\end{equation}
\begin{equation}
\bar U(x) \equiv \bar u_{\pi^+}(x) = \bar d_{\pi^-}(x);~~~\bar D(x) \equiv
d_{\pi^+}(x) = u_{\pi^-}(x)~.
\label{eq2}
\end{equation}
\begin{equation}
S(x) \equiv s_{\pi^+}(x) = \bar s_{\pi^-}(x);~~~\bar S(x) \equiv
\bar s_{\pi^+}(x) = s_{\pi^-}(x)~.
\label{eq3}
\end{equation}
\begin{equation}
G(x) \equiv  g_{\pi^+}(x) = g_{\pi^-}(x)~.
\label{eq4}
\end{equation}

The requirements of charge-conjugation (C) and charge symmetry (CS)
invariance would significantly reduce the number of independent parton
distributions of pion. 
As shown in Eq.~(\ref{eq1}),
C symmetry leads to $u_{\pi^+}(x) = \bar u_{\pi^-}(x)$.  Invariance
under the rotation in the isospin space by 180$^\circ$, i.e., 
CS invariance~\cite{Tim},
would give $\bar u_{\pi^-}(x) = \bar d_{\pi^+}(x)$. Therefore, invariance
under the combined operations of C and CS implies $U(x) = D(x)$. In a similar
fashion, it can be readily shown that $\bar U(x) = \bar D(x)$
and $S(x) = \bar S(x)$~\cite{BBP2021}.

Based on the framework of the statistical model, the four independent
pion parton distributions are expressed in the following
parametric forms:

\begin{equation}
xU(x)= xD(x) =\frac{A_U X_U x^{b_U}}{\exp[(x-X_U)/\bar x] + 1}
+\frac{\tilde A_{U} x^{\tilde b_{U}}}{\exp(x/\bar x) + 1}~.
\label{eq5}
\end{equation}

\begin{equation}
x\bar U(x) = x\bar D(x) =\frac{A_U (X_U)^{-1} x^{b_U}}{\exp[(x+X_U)/\bar x]
+ 1} +\frac{\tilde A_{U} x^{\tilde b_{U}}}{\exp(x/\bar x) + 1}~.
\label{eq6}
\end{equation}

\begin{equation}
x S(x) = x\bar S(x) = \frac{\tilde A_{U} x^{\tilde b_{U}}}{2[\exp(x/\bar x)
+ 1]}~.
\label{eq7}
\end{equation}

\begin{equation}
x G(x) = \frac {A_G x^{b_G}} {\exp(x/\bar x)-1},~~~b_G=1+\tilde b_U~.
\label{eq8}
\end{equation}

Following the formulation developed for proton's PDFs, the $x$ distributions
for fermions (quark and antiquark) have Fermi-Dirac parametric form, while
gluon has a Bose-Einstein $x$ distribution~\cite{BBS2002,BBS2005}.
The two terms for
$xU(x)$ and $x\bar U(x)$ in Eqs. (\ref{eq5}) and (\ref{eq6}) refer to
the nondiffractive
and diffractive contribution, respectively~\cite{BBS2002,BBS2005}.
As shown in the analysis of proton PDFs in the statistical
model~\cite{BBS2005}, the
presence of the diffractive term is important for describing the data
at the low $x$ region. 

A key feature of the statistical model is
that the chemical potential, $X_U$, for the quark distribution $U(x)$ becomes
$-X_U$ for the antiquark distribution $\bar U(x)$. The parameter $\bar x$
plays the role of the effective ``temperature."
For the strange-quark distribution $S(x)$, the requirement that $S(x)$ and
$\bar S(x)$ have identical $x$ distribution implies that the chemical
potential in the
nondiffractive term must vanish. Hence, the nondiffractive
and diffractive terms for $S(x)$ have the same parametric form,
and we make
the simple assumption that $S(x)$ is equal to half of the diffractive part
of  $\bar U(x)$ due to the heavier strange quark mass.
The expression $b_G = 1 + \tilde b_U$ in Eq.~(\ref{eq8}) has the 
interesting consequence that $G(x)$ has
an identical $x$ dependence as the diffractive part of the quark distributions
when $x \to 0$. The dominance of the gluon and sea-quark distributions at
$x \to 0$ and the strong interplay between them make this 
a very reasonable assumption.

Equations (\ref{eq5})$-$(\ref{eq8}) contain a total of 7 parameters, namely,
$A_U$, $X_U$, $b_U$,
$\bar x$, $\tilde A_U$, $\tilde b_U$, and $A_G$. These parameters are
also constrained by two sum rules, namely, the valence-quark number sum rule
and the momentum sum rule:

\begin{eqnarray}
 & \int_0^1 [U(x) - \bar U(x)]~dx =1~, \nonumber \\
 & \int_0^1 x [2 U(x) + 2 \bar U(x) + 2 S(x) + G(x)]~dx = 1~.
\label{eq:eq9}
\end{eqnarray}

\begin{figure}[tb]
\begin{center}
\includegraphics[width=8.0cm]{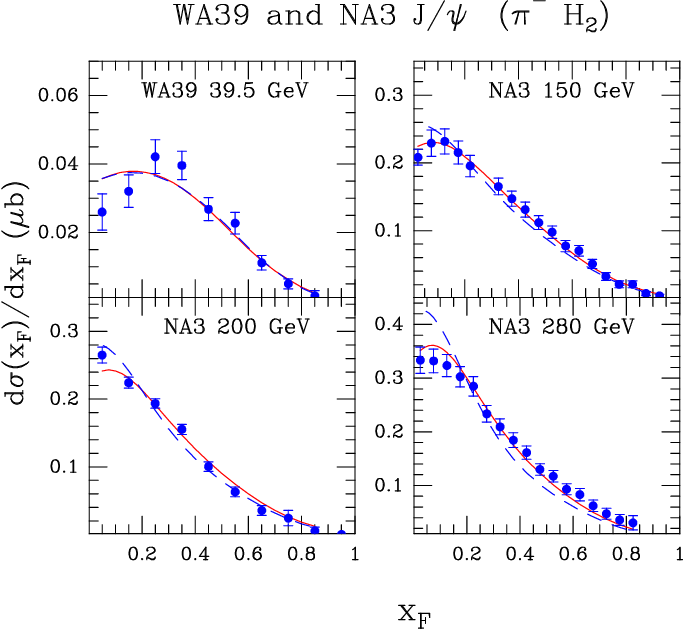}
\end{center}
\caption[*]{\baselineskip 1pt     
$J/\psi$ production data from the WA39 and NA3 experiment
$\pi^-  H_2$ at $P_{lab}^{\pi} = 39.5, 150, 200, 280~\mbox{GeV}$ 
for $d\sigma/dx_F$ (blue circles)
compared with the calculation of the BBP pion PDFs~\cite{BBP2021}
(dashed blue curves).
The solid red curves are results using the new pion PDFs.}
\label{fi1}
\end{figure}

\section{Global fit procedure}

In order to obtain the parameters for pion PDFs according to
the parametrizations listed in Eqs. (7) $-$ (10), we have 
fitted both the Drell-Yan and the $J/\Psi$ production data.
For the Drell-Yan data, we have performed next-to-leading-order (NLO)
QCD calculation to fit $\pi^-$-induced dimuon production
data on tungsten targets from E615 at 252 GeV~\cite{e615},
E326 at 225 GeV~\cite{e326}, and NA10 at 194 GeV and
286 GeV~\cite{NA10}. Detailed expressions for the NLO Drell-Yan cross
sections were presented in~\cite{BS2019}. The nucleon PDFs
used in the calculation were taken from the BS15 PDFs~\cite{BS2015},
obtained from a global fit to existing data in the
framework of the statistical model. The QCD evolution was performed using 
the HOPPET program~\cite{Salam2009}, and the CERN MINUIT 
program~\cite{James1994} was utilized for the $\chi^2$ 
minimization.
Since the Drell-Yan data in this analysis were all collected using
nuclear targets (tungsten), it is necessary to take into account the
nuclear modification of the nucleon PDFs, described in~\cite{BBP2021}. 

For the analysis of the $J/\Psi$ production data, we follow the 
recent study~\cite{Chang2021} on 
the comparison between pion-induced 
$J/\psi$ production data with theoretical model calculations
using the non-relativistic QCD (NRQCD)~\cite{NRQCD} approach. 
The NRQCD approach is
based on the factorization of the heavy-quark $Q\bar{Q}$ pair production
and its subsequent hadronization. The production of the $Q\bar{Q}$
pair involves short-distance partonic interaction,
calculated using perturbative QCD. The probability of 
a $Q\bar{Q}$ pair hadronizing into a quarkonium bound state 
is described by the long-distance matrix
elements (LDMEs). The LDMEs, assumed to be universal, are 
determined from the experimental data~\cite{Beneke:1996tk,Chang2021}.

We briefly describe the NRQCD framework used in this study as
formulated in Ref.~\cite{Beneke:1996tk,Vogt:1999dw}.
The differential cross section with respect to Feynman $x$
($x_F$), $d\sigma/dx_F$, for a charmonium state $H$ ($H$ = $J/\psi$,
$\psi(2S)$, or $\chi_{cJ}$) from the $hN$ collisions, where $h$ is 
the beam hadron and $N$ the target
nucleon, is~\cite{Vogt:1999dw}
\begin{align}
\label{eq:eq1}
\frac{d\sigma^{H}}{dx_F}& = & \sum\limits_{i,j=q, \bar{q},
  G} \int_{0} ^{1} dx_{1} dx_{2} \delta(x_F - x_1 + x_2) \nonumber \\
  & \times & f^{h}_{i}(x_1, \mu_{F}) f^{N}_{j}(x_2, \mu_{F}) \hat{\sigma}[ij 
\rightarrow H], 
\end{align}
where $i$ and $j$ label the type of interacting partons (gluons,
quarks and antiquarks), $f^{h}$ and $f^{N}$ are the
incoming hadron and the target nucleon parton distribution functions,
evaluated for their respective Bjorken-$x$ values, $x_1$ and $x_2$ at the
factorization scale $\mu_F$. $\hat \sigma$ is given as

\begin{align}
\hat{\sigma}[ij \rightarrow H] & = &  \sum\limits_{n} C^{ij}_{c \bar{c} [n]} 
(x_1 P_{h} , x_2 P_{N} , \mu_{F},
\mu_{R}, m_c) \nonumber \\
 & \times & \langle \mathcal{O}_{n}^{H}[^{2S+1}L_{J}] \rangle ,
\end{align}
where $C^{ij}_{c \bar{c} [n]}$ denotes the hard-process cross section
for producing a $c \bar{c}$ pair with
color ($n$), spin ($S$), orbital angular momentum ($L$) and total
angular momentum ($J$). The hadronization probability is specified
by the LDMEs, $\langle \mathcal{O}_{n}^{H}[^{2S+1}L_{J}] \rangle$.
Here $m_c$ and $M_{c \bar{c}}$ are the charm
quark and $c \bar{c}$ pair masses, and $\mu_R$ is the renormalization
scale.

\begin{figure}[tb]
\begin{center}
\includegraphics[height=6.8cm]{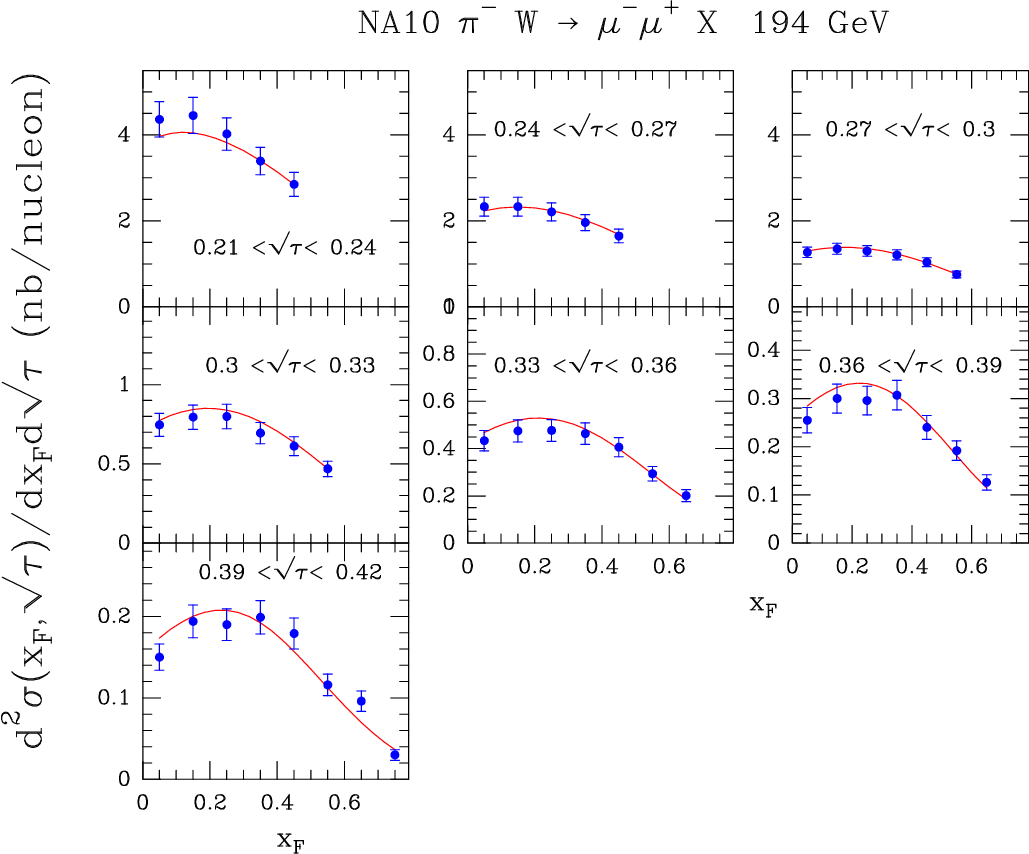}
\end{center}
\caption[*]{\baselineskip 1pt
Drell-Yan data from the NA10 experiment $\pi^- W$ at $P_{lab}^{\pi}
= 194~\mbox{GeV}$ \cite{NA10}.  $d^2\sigma/d\sqrt{\tau}dx_F$ versus $x_F$ for
several $\sqrt{\tau}$ intervals, where $\tau=m^2/s$, 
are compared with the results of the
new fit (solid curves).} 
\label{fi3}
\end{figure}

The LDMEs used in the NRQCD calculation were taken from a recent
study~\cite{Chang2021}, which extracts these matrix elements by
performed a fit to the energy dependence of the $x_F$-integrated
$J/\Psi$ production cross sections induced by proton and pion beams at
fixed-target energies. Several sets of the LDMEs were obtained in this
work, and we select the ``Fit-2" solution for the
LDMEs~\cite{Chang2021}. Using this set of LDMEs, the direct production
cross sections of $J/\Psi$, $\Psi(2S)$ and the three $\chi_{cJ}$
states are calculated using Eq.~(\ref{eq:eq1}). Furthermore, both the
direct production of $J/\psi$ and the feed-down from hadronic decays
of $\psi(2S)$ and radiative decays of three $\chi_{cJ}$ states have
been included for calculating the total $J/\psi$ cross section. More
details on the NRQCD calculation can be found in~\cite{Chang2021}. We
found that the results of the present analysis are not sensitive to
the choice of the specific LDME set obtained in~\cite{Chang2021},
since they were all constrained by the same $J/\Psi$ production data.

While there exist a significant number of measurements for pion-induced
$J/\Psi$ production cross sections as tabulated in~\cite{Chang:2020rdy}, 
a large fraction
of these data were collected using nuclear targets. Since large nuclear
effects for $J/\Psi$ production were found with both the proton and pion
beams, we only select $\pi^- + p \to J/\Psi$ data in our global fit. This
eliminates the uncertainties associated with the nuclear effects in
$J/\Psi$ production. The two $\pi^- + p \to J/\Psi$ experiments are the
CERN WA39 experiment at 39.5 GeV~\cite{WA39}, and the CERN NA3 experiment
at 150, 200, and 280 GeV~\cite{NA3,NA3_Thesis}.

The CERN WA39 Collaboration measured the $J/\Psi$~ production cross
section with 39.5 GeV hadron beams~\cite{WA39}. Data for the
liquid H$_2$ target were taken with negative and
positive hadron beams ($\pi^\pm$, $K^\pm$, $p$ and $\bar p$). 
The differential cross sections in $x_F$
for $\pi^- + p \to J/\Psi$
cover the region $0.05 \leq x_F \leq 0.85$. 
The normalization uncertainty on the cross sections is 15\%.

\begin{figure}[tb]
\begin{center}
\includegraphics[height=6.8cm]{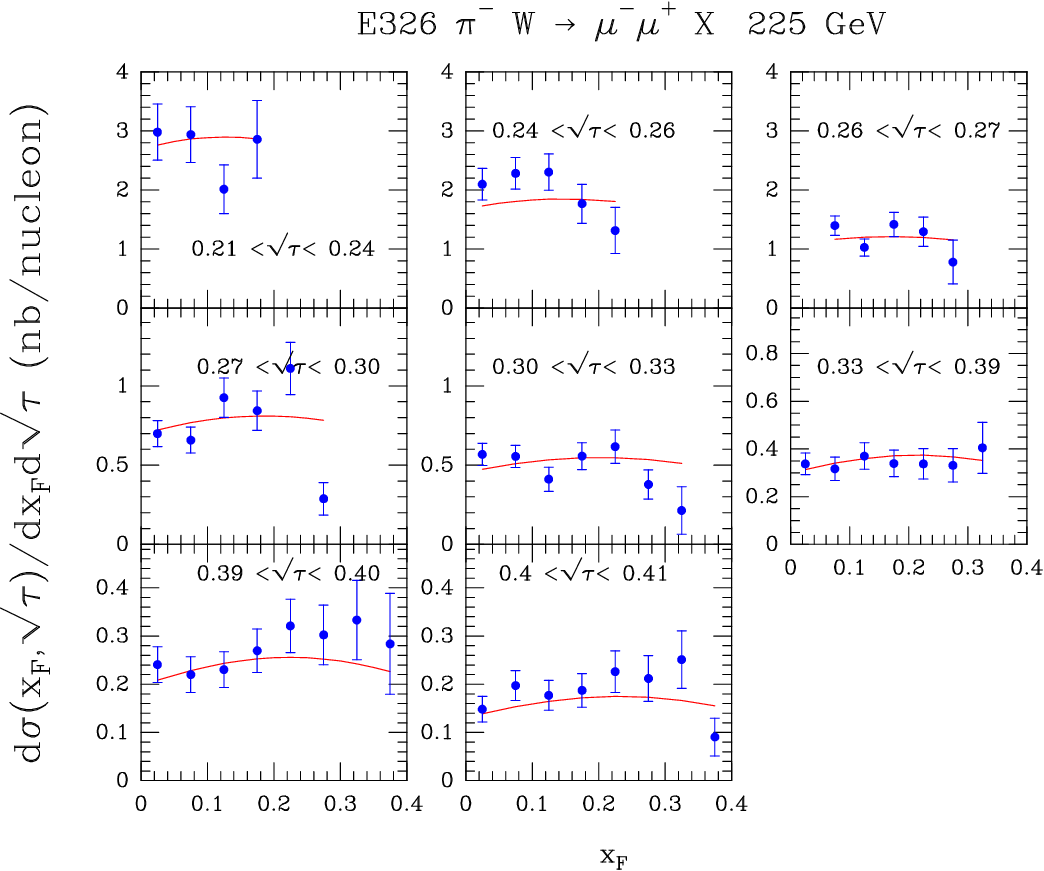}
\end{center}
\caption[*]{\baselineskip 1pt
Drell-Yan data from the E326 experiment $\pi^- W$ at $P_{lab}^{\pi}
= 225~\mbox{GeV}$ \cite{e326}.  $d^2\sigma/d\sqrt{\tau}dx_F$ versus $x_F$ for
several $\sqrt{\tau}$ intervals
are compared with the results of the
new fit (solid curves).} 
\label{fi4}
\end{figure}

The CERN NA3 experiment~\cite{NA3}, has 
the largest pion-induced $J/\Psi$~ production
statistics available to date. Data were collected at three different
incident momenta, 150, 200, and 280 GeV.
For all three beam energies, the cross sections have a
normalization uncertainty of 13\%. Although the numerical values
of the cross sections were not listed in the NA3 publication, they can 
be retrieved from the figures in the published 
paper~\cite{NA3} and an unpublished thesis~\cite{NA3_Thesis}.

\section{Results of the global fit}

Before presenting the results of the global fit, it is instructive to
compare the $J/\Psi$ production data with calculations using the recent
pion PDFs obtained in the statistical model~\cite{BBP2021}. These pion PDFs
were capable of reproducing the pion-induced Drell-Yan data very well.
Since these Drell-Yan data are mostly sensitive to the valence-quark
distributions, the gluon distribution in pion is only loosely
constrained from the momentum sum rule~\cite{BBP2021}. Figure~\ref{fi1}
shows the comparison between the data and the NRQCD calculation. While the
agreement between data and calculation is acceptable for the WA39 data
at 39.5 GeV ($\chi^2/ndp = 1.16$, where ndp is the number of data
points), a much larger value of 
$\chi^2/ndp = 3.11$ is obtained for the NA3 data at higher pion beam
energies. The $J/\Psi$ production at 39.5 GeV is dominated
by the $q \bar q$ annihilation process and is only sensitive to
the valence-quark distribution in the 
pion~\cite{Chang:2020rdy,Chang2021}. As the valence-quark distribution
is rather well determined from the BBP 
fit to the Drell-Yan data~\cite{BBP2021},
it is reassuring that the $J/\Psi$ data at 39.5 GeV is well described
by the NRQCD calculation. In contrast, the gluon-gluon fusion process
has increasing important contributions as the pion beam 
energy increases~\cite{Chang:2020rdy,Chang2021}.
The poor agreement between the calculation and the NA3 data at 
higher beam energies clearly indicates that the gluon distribution obtained
from the fit to Drell-Yan data alone is not adequate. 
Combining the Drell-Yan
and $J/\Psi$ data in the global fit, as shown later, could 
lead to an improved extraction of both the gluon and the valence-quark 
distributions in the pion. 

\begin{figure}[tb]
\begin{center}
\includegraphics[height=6.8cm]{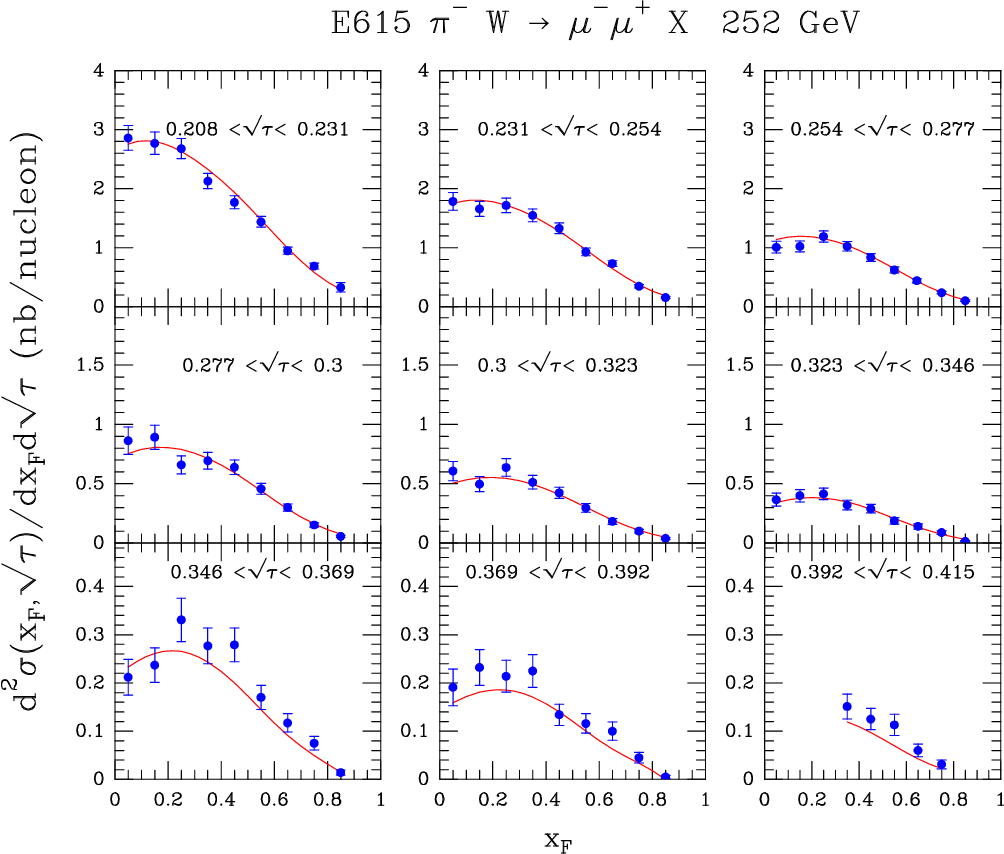}
\end{center}
\caption[*]{\baselineskip 1pt
Drell-Yan data from the E615 experiment
$\pi^- W$ at $P_{lab}^{\pi} = 252~\mbox{GeV}$ \cite{e615}.
$d^2\sigma/d\sqrt{\tau}dx_F$ versus $x_F$ for several $\sqrt{\tau}$
intervals are compared with the results of the
new fit (solid curves).}  
\label{fi5}
\end{figure}

Following the procedure described in Secs. II and III, the best-fit
values for the various parameters in the statistical model are
obtained. Table I lists the number of data points and the values of
$\chi^2$ for the best fit to these datasets. Note that the
normalizations for the absolute cross sections from various
experiments contain systematic uncertainties on the order of $\sim$ 15
percents. In the global fit, the normalizations for various datasets
are allowed to vary, as listed in Table I, in order to achieve
improved consistency among various datasets. The result of the
calculation is multiplied by the K factor when compared with the
data. We find that the K factors for the fit to Drell-Yan data are
within 10\% of unity, consistent with the normalization uncertainties
of the experiments. For the $J/\Psi$ data, the values of K factors are
also found to be close to 1 for the NA3 data at three beam
energies. However, the K factor for the WA39 experiment at 39 GeV is
quite small, 0.63.  A very similar finding was reported in the
comparison between NRQCD calculation using previous pion PDFs with the
$J/\Psi$ production data~\cite{Chang:2020rdy}. This might reflect a
slight underestimate of the normalization uncertainties from this
experiment. It could also suggest that further improvement in the
NRQCD model is warranted.

\begin{table}[htbp]
\caption {Values of the K factor, $\chi^2$ and $\chi^2/ndp$ (where $ndp$ is
the number of data points) for each
dataset obtained from a global fit.
P is the beam momentum, K the normalization
factor to be multiplied by the calculation for the Drell-Yan
and $J/\Psi$ cross sections,
$N_{data}$ the number of data points, $\chi^2$ and the $\chi^2/ndp$.
}
\begin{center}
\begin{tabular}{ c c c c c c}
\hline
\hline
Experiment   &P(GeV) &K  &$N_{data}$ & $\chi^2$ & $\chi^2/ndp$ \\
\hline\raisebox{0pt}[12pt][6pt]
E615    & 252    & 0.94   & 91  & 125 & 1.37 \\[4pt]
E326    & 225    & 1.07   & 50  & 77  & 1.53 \\[4pt]
NA10    & 286    & 1.12   & 23  & 10 & 0.44  \\[4pt]
NA10    & 194    & 1.14   & 44  & 22  &  0.49\\[4pt]
WA39 $J/\psi$ & 39   &  0.63   & 9    & 11  & 1.22 \\[4pt]
NA3 $J/\psi$ & 150   & 1.13    & 18  & 9.6 &  0.53 \\[4pt]
NA3 $J/\psi$ & 200   & 0.92   & 9   & 14  & 1.56 \\[4pt]    
NA3 $J/\psi$ & 280   & 1.00    & 17  &17 &  1.0\\[4pt]
\hline\raisebox{0pt}[12pt][6pt]
Total                        & &          & 261  &  284 & 1.09\\[4pt]
\hline
\hline
\end{tabular}
\label{table1}
\end{center}
\end{table} 

\begin{figure}[tb]
\begin{center}
\includegraphics[height=5.5cm]{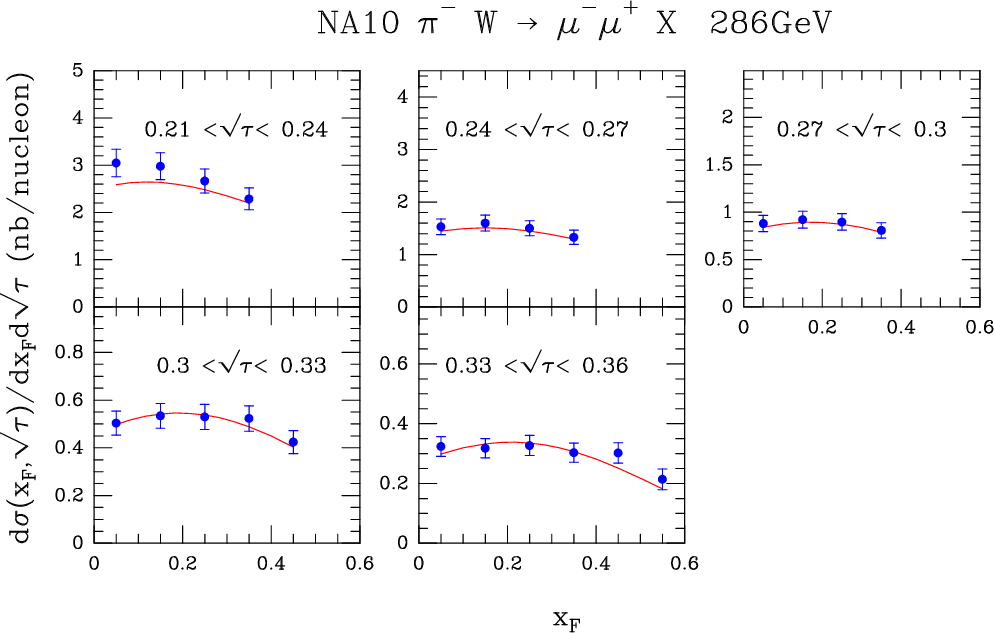}
\end{center}
\caption[*]{\baselineskip 1pt
Drell-Yan data from the NA10 experiment $\pi^- W$ at
$P_{lab}^{\pi} = 286~\mbox{GeV}$ \cite{NA10}.
$d^2\sigma/d\sqrt{\tau}dx_F$ versus $x_F$ for several
$\sqrt{\tau}$ intervals are compared with the results of the
new fit (solid curves).}  
\label{fi6}
\end{figure}

The $\chi^2$ values listed in Table~\ref{table1} shows that 
a simultaneous fit to the Drell-Yan and $J/\Psi$ data in the statistical
model with a small number of parameters can be 
obtained. In particular, the $\chi^2/ndp$ 
values are significantly smaller than that obtained with the previous
pion PDFs~\cite{BBP2021}. Figure~\ref{fi1} shows the good agreement 
between the $J/\Psi$
production data and the calculation with the new pion PDFs.
In Figs.~$2-5$, the fits to the Drell-Yan data using
the current result on pion's PDFs are displayed.
We note that the Drell-Yan data remain well described by
the statistical model, while much better agreement with the
 $J/\Psi$ data is obtained with the new set of pion PDFs.

The best-fit parameters of the pion PDFs, obtained at an initial 
scale $Q^2_0 = 1$ GeV$^2$, are

\begin{align}
A_U  &=  1.11 \pm 0.05 & b_U & = 0.64 \pm 0.02  \nonumber \\
X_U  &=  0.72 \pm 0.01   & \bar x & = 0.119\pm 0.002  \nonumber \\
\tilde A_U  &= 2.68 \pm 0.16 & \tilde b_U & = b_G - 1. \nonumber \\
A_G  &=  48.5 \pm 1.3 & b_G & = 1.88 \pm 0.04.
\label{eq14}
\end{align}

It is worth noting that the temperature, $\bar x = 0.119$, found for
pion is very close
to that obtained for proton, $\bar x = 0.090$~\cite{BS2015}, indicating a
common feature for the statistical description for baryons and mesons.
On the other hand, the chemical potential of the valence quark for pion,
$X_U = 0.72$, is significantly large than that for proton,
$X_U^+ = 0.475, X_U^- = 0.307$~\cite{BS2015}.

Figure 6 displays $xU(x)$, $x \bar U(x)$, $x S(x) = x \bar S(x)$, and
$x G(x)$ at $Q^2 = 10$ GeV$^2$ obtained in the present
analysis. Comparisons with the distributions from the previous
analysis in the statistical model~\cite{BBP2021} and global fits of
SMRS~\cite{Sutton:1991ay} and the more recent JAM~\cite{JAM} are also
shown in Fig. 6.  The shape and magnitude of the pion PDFs obtained in
the statistical model analysis are different from that of SMRS and
JAM. This reflects the very different parametric forms for the PDFs in
the statistical model compared with that of the conventional global
fits. We note that the gluon distribution of the new pion PDFs from
this analysis is significantly larger than that of the JAM PDFs for
the $x > 0.1$ region. Since the $J/\Psi$ production data from NA3 is
sensitive to the gluon distributions at the large $x$ region, an
improved determination of the gluon content in the pion is expected
for the $x > 0.1$ region. Table II lists the momentum fractions
carried by the quark, antiquark, and gluon in the pion for the new
PDFs obtained in this work at $Q^2 = 10$ GeV$^2$. The corresponding
momentum fractions for other pion PDFs are also shown for comparison.

\begin{figure}[hbp]
\begin{center}
\includegraphics[width=5.5cm]{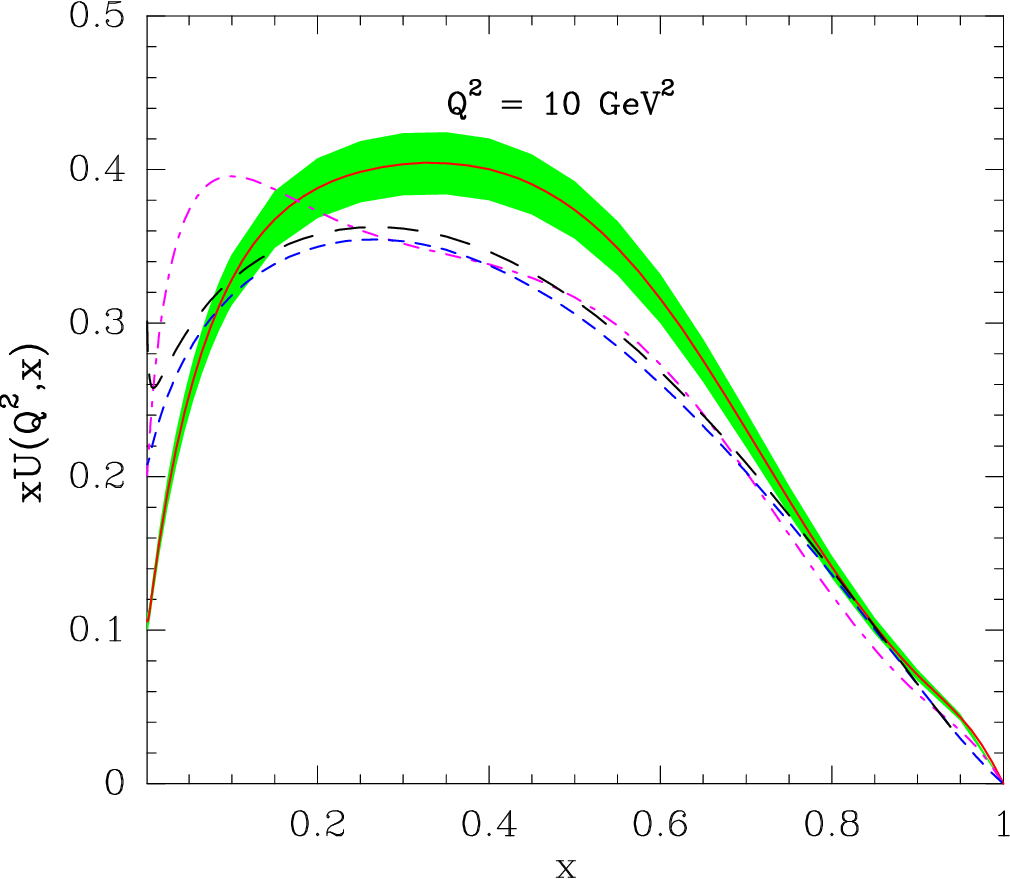}
\includegraphics[width=5.5cm]{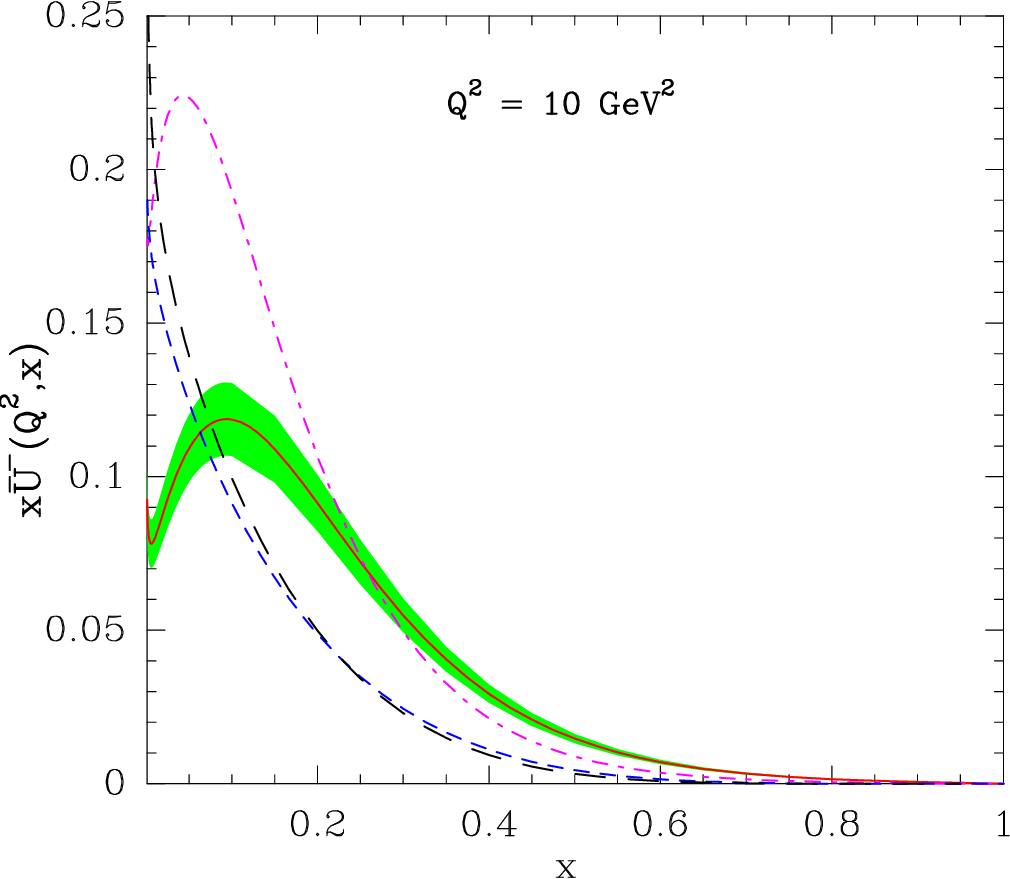}
\qquad
\includegraphics[width=5.5cm]{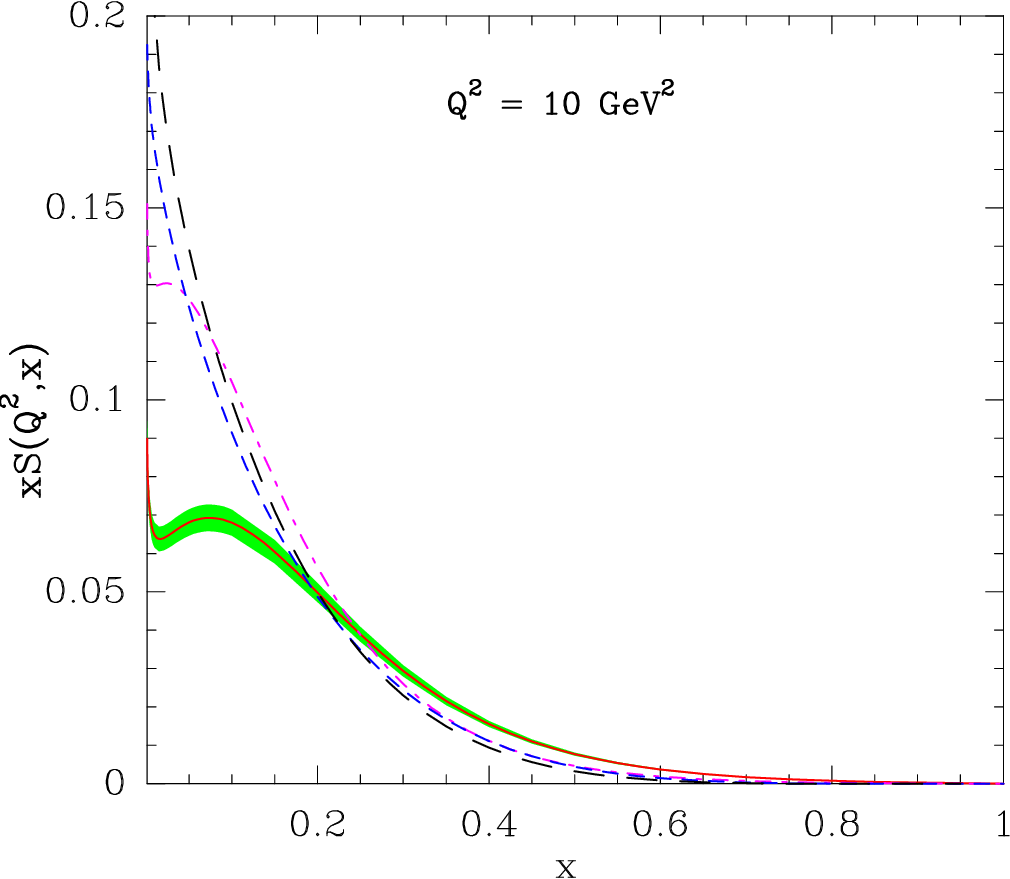}
\includegraphics[width=5.5cm]{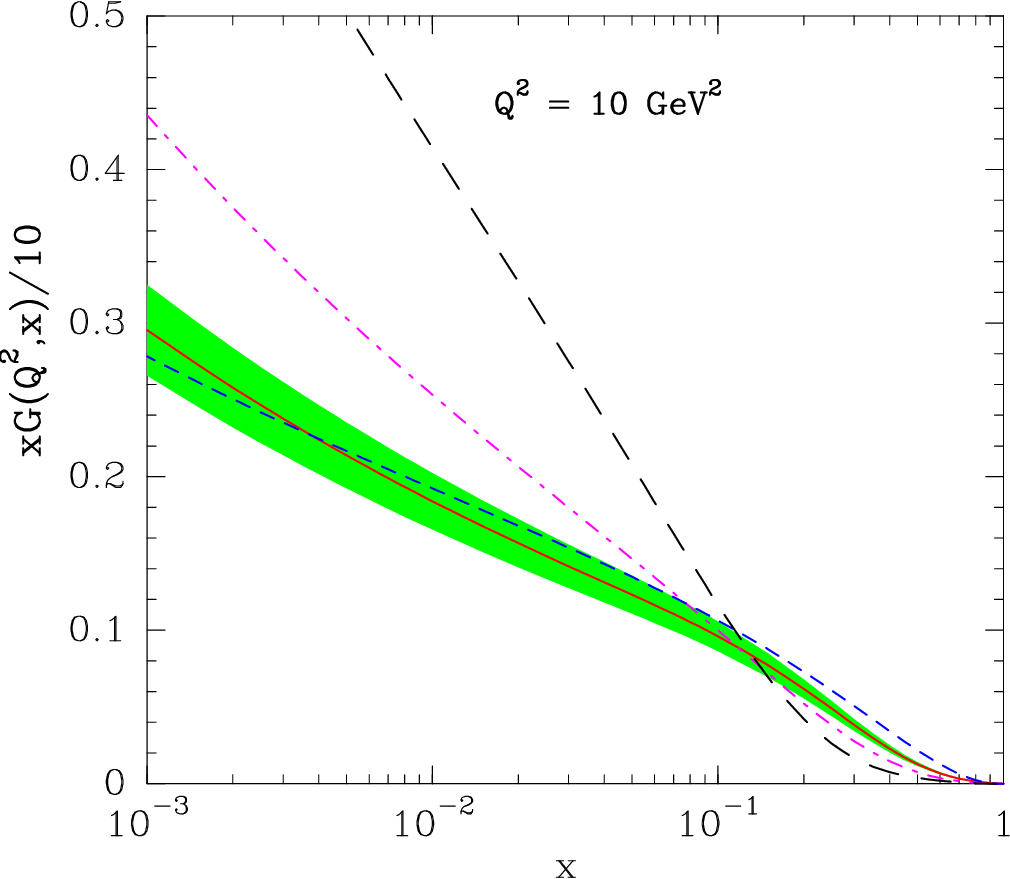}
\end{center}
\caption[*]{\baselineskip 1pt     
Different $\pi^-$ parton distributions, $xU$, $x\bar U$, $xS$, and $xG$,
versus $x$, after NLO QCD
evolution at $Q^2=10~\mbox{GeV}^2$. Present statistical model (solid, red),
BBP PDF from Ref.~\cite{BBP2021} (dotted-dashed, violet),
SMRS PDFs from Ref.~\cite{Sutton:1991ay} (dashed, blue),
JAM PDFs from Ref.~\cite{JAM2021} (long-dashed, black) are shown.
For SMRS  $x\bar S(x) = x\bar U(x)$.}
\label{PDFpi}
\end{figure}

\begin{table}[tbp]   
\caption {Momentum fractions of valence quarks, sea quarks and gluons
  of various pion PDFs for $\pi^-$ at the scale $Q^2$= 10
  GeV$^2$.\\ $^{a}$Uncertainties estimated from the member PDF sets.
BCP refers to the present work.}
\begin{center}
\begin{tabular}{|c|c|c|c|}
\hline
\hline
PDF
& $\int_0^1 x\bar{u}_{val}(x) dx$ & $\int_0^1 x\bar{u}_{sea}(x) dx$ & $\int_0^1 xG(x) dx$ \\
\hline
OW & 0.176 & 0.026 & 0.488 \\
ABFKW & 0.178 & 0.026 & 0.468 \\
SMRS & 0.219 & 0.026 & 0.395 \\
GRV & 0.179 & 0.020 & 0.513 \\
JAM$^{a}$ & $0.222 \pm 0.005$ & $0.028 \pm 0.002$ & $0.367 \pm 0.016$ \\
xFitter$^{a}$ & $0.230 \pm 0.008$ & $0.036 \pm 0.014$ & $0.309 \pm 0.065$ \\
BCP & 0.277 $\pm 0.003$ & 0.035 $\pm 0.002$ & 0.326 $\pm 0.015$ \\ 
\hline
\hline
\end{tabular}
\end{center}
\label{tab:xmoment}
\end{table}

\section{Conclusion and future prospects}

We have performed a new analysis to extract pion's PDFs in the
statistical model via a global fit to existing $\pi^-$-induced
Drell-Yan data as well as the $\pi^- + p \to J/\Psi$ data. Using a
parametrization of pion PDFs containing only a few parameters, a good
description of the Drell-Yan and $J/\Psi$ production data can be
obtained. The $J/\Psi$ production data at the lowest pion beam energy,
39.5 GeV, are sensitive to the pion valence-quark distribution,
analogous to the Drell-Yan data. At higher beam energies, the
pion-induced $J/\Psi$ production data probe the gluon distribution of
pion. A combined analysis of the Drell-Yan and $J/\Psi$ production
data has provided an improved determination of both the valence-quark
and the gluon distributions of pion compared with earlier studies.

From the new analysis, we confirm the previous result~\cite{BBP2021}
that the statistical model approach gives very similar values for the
temperature parameters for proton and pion, suggesting the consistency
of this approach for different hadronic systems. A larger value of the
valence-quark chemical potential for pion than for the proton is also
found.  New pion-induced Drell-Yan and $J/\Psi$ production data
anticipated from COMPASS and AMBER~\cite{AMBER} would provide further
tests of the pion PDFs obtained in the statistical approach.

The finding that both the Drell-Yan and $J/\Psi$ production data can
be well described with the statistical model approach suggests that
this approach could be extended to extract the kaon
PDFs~\cite{Peng2017}.  While only a single low-statistics measurement
of the kaon-induced Drell-Yan cross section is
available~\cite{NA3_KP}, there exist additional kaon-induced $J/\Psi$
production data at 39.5 GeV~\cite{WA39} and 200 GeV~\cite{NA3}. A
combined analysis of these kaon-induced Drell-Yan and $J/\Psi$
production might lead to a first extraction of the valence-quark and
gluon distributions of kaon. The proposed RF separated kaon
beam~\cite{AMBER} at CERN would be extremely valuable for measuring
the kaon PDFs in the future.

We acknowledge interesting discussions with Prof. F. Buccella during
the course of this work. This work was supported in part by the
U.S. National Science Foundation Grant No. PHY-1812377 and the
Ministry of Science and Technology in Taiwan, ROC.

\end{document}